\documentclass[%
reprint,
 amsmath,amssymb,
 aps,
prl
]{revtex4-1}

\usepackage{graphicx}% Include figure files
\usepackage{dcolumn}% Align table columns on decimal point
\usepackage{bm}% bold math

\usepackage[usenames]{xcolor}

\begin{document}

%\preprint{INT-PUB-18-055}

\title{Gravimeter search for compact dark matter objects moving in the Earth}


\author{C. J. Horowitz}
\email{horowit@indiana.edu}
\affiliation{Center for Exploration of Energy and Matter and Department of Physics, Indiana University, Bloomington, IN 47405, USA}

\author{R. Widmer-Schnidrig}
\email{widmer@gis.uni-stuttgart.de}
\affiliation{Black Forest Observatory (BFO), Heubach 206, D-77709 Wolfach, Germany} \affiliation{Institute of Geodesy, Stuttgart University, Stuttgart, Germany}

\date{\today}% It is always \today, today,

\begin{abstract}
Dark matter could be composed of compact dark objects (CDOs).  These objects may interact very weakly with normal matter and could move freely {\it inside} the Earth.  A CDO moving in the inner core of the Earth will have an orbital period near 55 min and produce a time dependent signal in a gravimeter.  Data from superconducting gravimeters rule out such objects moving inside the Earth unless their mass $m_D$ and or orbital radius $a$ are very small so that $m_D\, a < 1.2\times 10^{-13}M_\oplus R_\oplus$.  Here $M_\oplus$ and $R_\oplus$ are the mass and radius of the Earth.
\end{abstract}

\pacs{Valid PACS appear here}% PACS, the Physics and Astronomy
\maketitle

%\tableofcontents

%\section{\label{sec:intro}Introduction}
Many dark matter direct detection experiments have not yet seen a clear signal.  Limits from most of these experiments can be avoided if dark matter is concentrated into macroscopic objects.  Dark matter, or one component of it, could be composed of compact dark objects (CDOs).  These objects are assumed to have small non-gravitational interactions with normal matter and could be primordial black holes, see for example \cite{PhysRevD.95.083508}.  Some other possibilities or names for CDOs include Boson Stars \cite{Boson_stars}, Dark Blobs \cite{PhysRevD.98.115020}, asymmetric dark matter nuggets \cite{PhysRevD.97.036003}, Exotic Compact Objects \cite{Giudice_2016}, Ultra Compact Mini Halos (UCMH) \cite{PhysRevD.85.125027} made for example of axions \cite{Yang:2017cjm}, and Macros \cite{10.1093/mnras/stv774}.     Microlensing observations rule out most of dark matter being made of CDOs with masses between $10^{-11}M_\odot$ and $15M_\odot$ \cite{microlensing2019,Alcock:1995dm,Alcock:2000ph,Tisserand:2006zx,Paczynski:1986}.
In this paper, we focus on CDOs with masses between about $10^{-19}$ and $10^{-11}M_\odot$.  We assume the objects are not black holes (to avoid destroying the Earth) but otherwise try to minimize our assumptions about detailed CDO properties.  

Dark matter is known to have gravitational interactions.  Therefore, it is appealing to search for dark matter using gravity.   Compact dark objects can radiate detectable gravitational waves (GWs) \cite{Cardoso2019,Giudice_2016}.   The LIGO-Virgo collaboration searched for GWs from CDO binaries with masses in the range $0.2-1~M_\odot$ \cite{Abbott:2018oah}.   We explored GW signals from CDOs merging with neutron stars \cite{Horowitz_Reddy}.  In addition, we searched for GWs from CDOs orbiting inside the sun \cite{Horowitz:2019pru}, and ruled out close binaries with masses above $10^{-9}~M_\odot$.    

To probe CDO masses well below $10^{-9}~M_\odot$, we now consider CDOs moving around or {\it inside} the Earth.  It can be difficult to constrain such low mass objects with microlensing \cite{microlensing2019}, or femtolensing \cite{Katz_2018}, because of the small size of the lens compared to the background star or gamma ray burst.  Instead, nearby CDOs could produce detectable signals in gravimeters \cite{doi:10.1063/1.1150092} that measure the local acceleration due to gravity.  

Sensitive superconducting gravimeters have been deployed at several locations around the world \cite{IGETS}. They are used to observe a wide range of geophysical phenomena including Chandler wobble, solid Earth tides, post glacial rebound, seismic free oscillations and hydrology \citep{HindererCrossleyWarburton:07}.
In addition to geophysics, they have been used to search for a dependence of gravity on a hypothetical preferred reference frame \cite{1972ApJ...177..757W,1976ApJ...208..881W}, or the violation of Lorentz invariance \cite{PhysRevLett.119.201101,PhysRevD.97.024019}, as the Earth translates or rotates.  In addition, gravimeters have been used to search for oscillations of the Earth excited by gravitational waves \cite{PhysRevD.90.042005}.

If there are many CDOs moving through the inner solar system, it is possible that over the solar system's lifetime a three body interaction (such as a close encounter of a binary system with the Earth) or some other mechanism could lead to the capture of a CDO in orbit around, {\it or through}, the Earth.  For example, Neptune's moon Triton is thought to have been captured in this way \cite{triton}.   Although capture might be rare, it could greatly aid the detection of what otherwise are probably very difficult to observe objects.   As an alternative to relying on capture, one could search for unbound CDOs moving through the Earth, see for example \cite{Meteors}.  However for our mass range, such events are likely extremely rare.  

We assume that the unknown interactions between the CDO and earth matter are small enough so that the CDO can move through the Earth with only modest dissipation.  If so, this modest dissipation from dynamical friction \cite{Ostriker_1999,Kim_2010} and or additional weak non-gravitational interactions could cause the orbit to slowly decay so that today the CDO could be orbiting {\it inside} the Earth's inner core.   We note that the CDO will move subsonically, unless the radius of its orbit is nearly the radius of the Earth (or larger).    Dynamical friction, for subsonic motion in a gas, could lead to an orbital decay time of order $(T M_\oplus)/m_D$ \cite{Ostriker_1999}, where $T$ is the orbital period (see below), $M_\oplus$ is the mass of the Earth, and $m_D$ is the mass of the CDO.    For $m_D\approx 10^{-12}M_\oplus$ this decay time is of order $10^8$ years.

 \begin{figure}[ht]
\smallskip
\includegraphics[width=1.0\columnwidth]{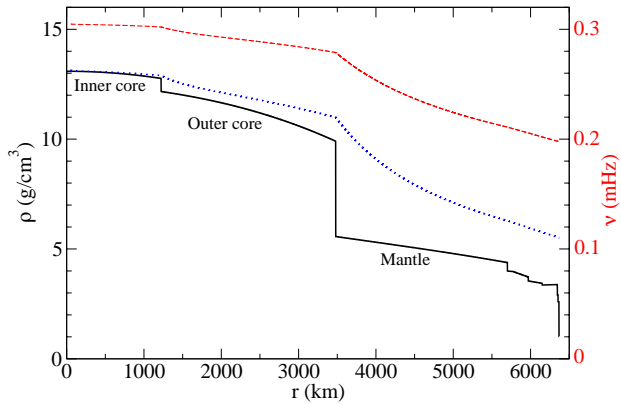}
 \caption{(Color online) 
Density of the Earth $\rho(r)$ versus radius $r$ (solid black line)\cite{DZIEWONSKI1981297}.  The dotted blue line shows the average density (of matter interior to $r$) $\bar\rho(r)$.  Finally the orbital frequency $\nu$ for a circular orbit inside the Earth of radius $r$ is shown as the dashed red line using the righthand scale.}
\label{Fig1}
\end{figure}

An object moving in a circular orbit, through an average density $\bar\rho$, will orbit with period $T$ and frequency $\nu$ given by Kepler's 3rd law,   
\begin{equation}
\nu=\frac{1}{T}=\bigl(\frac{1}{3\pi} G \bar\rho\Bigr)^{1/2}\, .
\label{omega}
\end{equation}
The density of the Earth $\rho(r)$ \cite{DZIEWONSKI1981297} is plotted in Fig. \ref{Fig1}  along with the average density $\bar\rho(r)$ of matter interior to radius $r$. The orbital frequency $\nu$ varies from $\approx0.3$ mHz for small $r$ to 0.2 mHz at the surface.
Near the center of the Earth $\bar\rho\approx \rho_c=13.1$ g/cm$^3$ and is nearly constant \cite{DZIEWONSKI1981297}. For a constant density, the orbits are ellipses with the center of the ellipse coincident with the center of the Earth and the period is independent of $r$.

Gravimeters on the surface of the Earth could be sensitive to CDOs moving in the inner core by looking for very small periodic changes in the local acceleration from gravity (little $g$) with period near $T=55$ min or frequencies near $\nu(\bar\rho=\rho_c)=\nu_0=0.305$ mHz (or somewhat smaller for larger radius orbits).  In general, we don't know $m_D$ or the radius of the orbit.  However, we know the (approximate) orbital frequency $\nu_0$ because we know the density profile inside the Earth. 

As a first example, consider a gravimeter at the north pole and a CDO that is oscillating along the Earth's rotation axis with time dependent position $x(t)=a\cos(\omega t)$.
Here $a$ is the amplitude of the motion and $\omega=2\pi\nu_0$.    The center of mass of the Earth will recoil so that its 
acceleration is,
$
{d^2X_\oplus(t)/dt^2}=\omega^2\, (m_D/M_\oplus)a \cos(\omega t)$.  We assume $m_D\ll M_\oplus$.
The gravimeter will have a time dependent reading for two reasons.  First, the meter is accelerating because it is on the (assumed rigid) Earth that is recoiling.  Second, the gravitational acceleration due to the CDO will change with time as the distance between the CDO and the gravimeter changes.  As we will see, both contributions are of order $(m_Da)/(M_\oplus R_E)$ times $g$.  Here the acceleration due to Earth's gravity is $g=GM_\oplus/R_\oplus^2$.

The gravitational acceleration $g_D$ at the gravimeter from the CDO is,
\begin{equation}
g_D=\frac{Gm_D}{\bigl(R_\oplus-a \cos(\omega t)\bigr)^2}\approx \frac{Gm_D}{R_\oplus^2}[1+2\frac{a}{R_\oplus}\cos(\omega t)]\, ,
\label{g_D}
\end{equation}
assuming $a\ll R_\oplus$.  The time dependent total gravimeter reading $\Delta g(t)$ is,
 \begin{equation}
 \Delta g(t)\approx\Bigl[\omega^2 a \frac{m_D}{M_\oplus} + 2 \frac{G m_D a}{R_\oplus^3} \Bigr]\cos(\omega t)\, .
 \label{Delta g}
 \end{equation}
 Using $\omega$ from Eq. \ref{omega} (with $\bar\rho\approx\rho_c$ given $a\ll R_\oplus$) and introducing the average density of the Earth $\bar\rho_\oplus=3M_\oplus/(4\pi R_\oplus^3)\approx 5.51$ g/cm$^3$, Eq. \ref{Delta g} can be written,
 \begin{equation}
 \frac{\Delta g(t)}{g}\approx(2+\frac{\rho_c}{\bar\rho_\oplus})(\frac{m_D a}{M_\oplus R_\oplus})\cos(\omega t)\, .
\label{deltag}
 \end{equation}
Thus the gravimeter reading oscillates at frequency $\nu_0=0.305$ mHz with fractional amplitude (compared to $g$) of order $(m_D a)/(M_\oplus R_\oplus)$.

We now consider the more general case where the CDO is in a circular orbit of radius $a$ ($\ll R_\oplus$) that is inclined by an angle $\Theta_I$ with respect to the plane of the equator.  Let the gravimeter be located at Latitude $\Theta_L$.  The time dependent gravimeter reading is now \cite{Suplemental_info},
\begin{equation}
    \frac{\Delta g(t)}{g}=(2+\frac{\rho_c}{\bar\rho_\oplus})
    \left(\frac{m_D a}{M_\oplus R_\oplus}\right)\Delta(t)\,
    \label{Eq.dgfull}
    \end{equation}
with $\Delta(t)=\delta_1\cos(\omega-\omega_\oplus)t+\delta_2\cos(\omega+\omega_\oplus)t+\delta_3\sin\omega t$.  Here $\delta_1=\cos\Theta_L\cos^2\Theta_I/2$, $\delta_2=\cos\Theta_L\sin^2\Theta_I/2$, and $\delta_3=\sin\Theta_L\sin\Theta_I$.  In addition to the original signal at angular frequency $\omega$, there are now signals at the rotational side band frequencies $\omega\pm\omega_\oplus$ with  $\omega_\oplus=2\pi/{\rm day}$.  This is because the gravimeter rotates with the Earth.   

Equation \ref{Eq.dgfull} for a circular orbit has a very similar form to Eq. \ref{deltag} for an extremely eccentric orbit.  Therefore, we do not expect the gravimeter signal to depend strongly on the eccentricity of the orbit.  Finally, the orientation of the orbit will slowly advance in time because the Earth's density is not constant \cite{Suplemental_info}.  However, we don't expect this to significantly modify the gravimeter signal.

We now analyze gravimeter data.  A number of superconducting gravimeters (SGs) have been deployed at various locations around the world.  Data from these devices has been archived by the Global Geodynamics Project (GGP, 1997-2015) and by the International Geodynamics and Earth Tide Service (IGETS, 2015-) \cite{IGETS,gravimeter_data}.  The sensor self-noise of these instruments improved in 2009 when the manufacturer increased the mass of the levitated proof mass from 4 to 17.2 grams \citep{RosatHinderer_2011}. At the Black Forest Observatory (BFO at $48.33^\circ$N, $8.33^\circ$E) in South-Western Germany the first of these new SGs was installed and because of its low noise level we will concentrate our analysis on data from that instrument. The analysis is complemented with data from the SG in Canberra (CB at $35.32^\circ$S, $149.01^\circ$E)
\begin{figure}[ht]
\smallskip
\includegraphics[width=1.0\columnwidth]{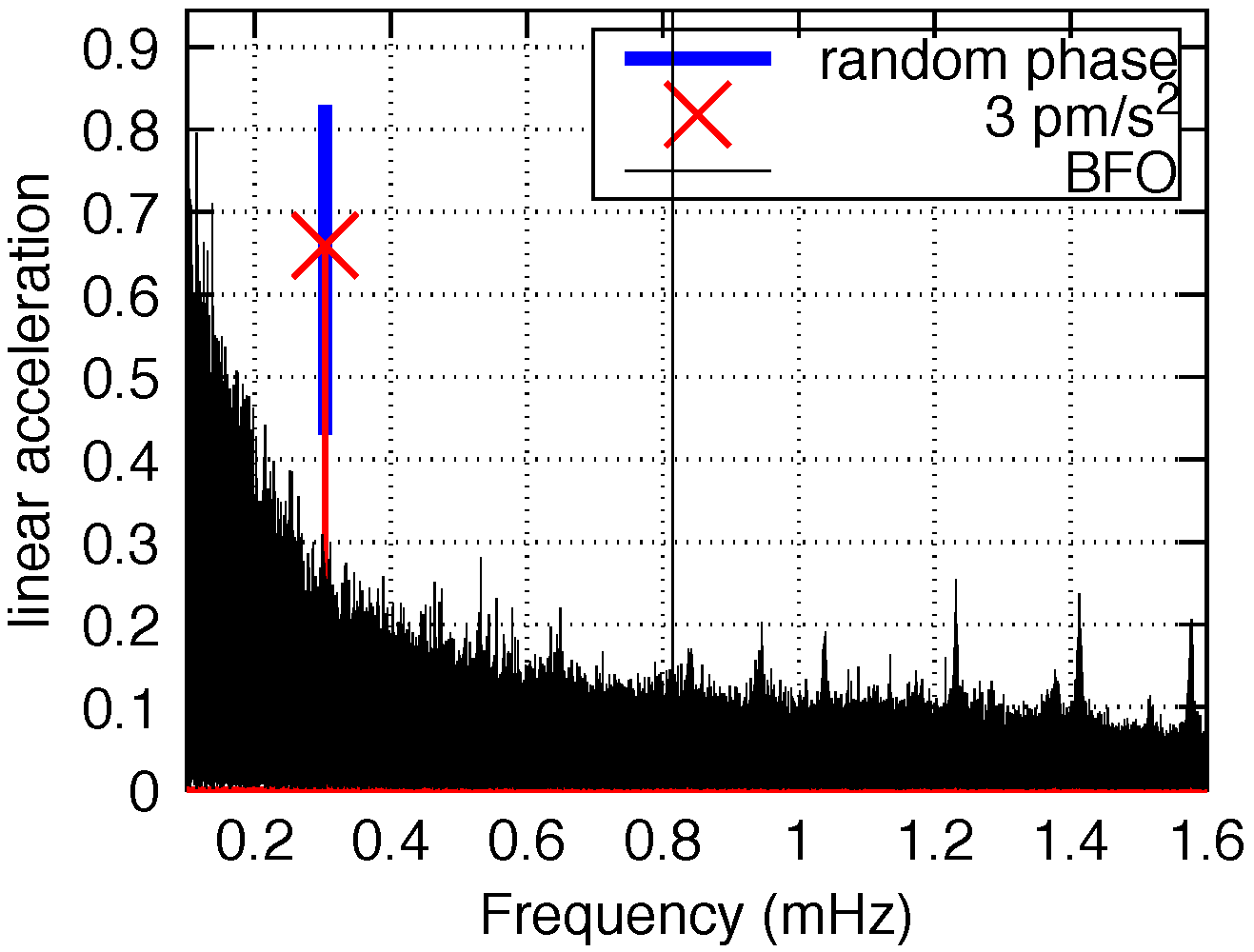}
 \caption{ 
(Color on line) Fourier amplitude spectrum of gravity residuals at station BFO (Black Forest Observatory) versus frequency, solid black. The time series starts on July 20, 2011 and is 6.7 years long. The gravity data have been corrected for the atmospheric pressure with an admittance of -3.0 nms$^{-2}$/hPa.  The red line with a cross shows calibration injections at a frequency of 0.303 mHz with an amplitude of 3 pm/s$^2$ and different phases shown by the blue band.}
\label{Fig2}
\end{figure}

We analyze about 10 years of data from each instrument. The gravimeter
data contains signals from numerous phenomena including tides, earthquakes and atmospheric processes. The frequency band of interest here - 0.2 to 0.3 mHz - is above the tides but overlaps with the lowest order seismic free oscillation. The largest signal in this band is from Newtonian attraction of variable air masses in the atmosphere above the sensor \cite{HindererCrossleyWarburton:07, ZuernWielandt_2007}. Since this is a well known effect all gravimeters are also equipped with a continuously recording barometer.  For the simple model of a horizontally layered atmosphere over a rigid half space the admittance between a pressure perturbation and the resulting gravity perturbation is: $\Delta g/\Delta p=-2\pi G/g = -4.27$ nms$^{-2}$/hPa.
This admittance is reduced by two smaller but related effects that both have opposite sign to the Newtonian attraction effect. The indentation of the Earth’s crust by the barometric load leads to (1) an inertial downward acceleration and (2) a vertical motion of the gravimeter in a gradient field.
Since we don't know the rigidity of the Earth's crust at the site of the gravimeter and since we anticipate that the admittance also exhibits some frequency dependence \cite{WarburtonGoodkind:1977} we estimate the gravity-pressure admittence with a  one parameter least squares regression.
We obtain empirical admittances of -3.0 and -2.9 nms$^{-2}$/hPa for BFO and CB, respectively. 
To remove the atmosphere induced signal we use these admittances as scale factors and subtract the locally recorded barometric pressure from the raw gravity data. 
We find that in our frequency band the pressure correction is very efficient.  In fact the atmosphere accounts for 60\% of the raw gravity signal while other sources account for less than 40\%. 

\begin{figure}[ht]
\smallskip
\includegraphics[width=1.0\columnwidth]{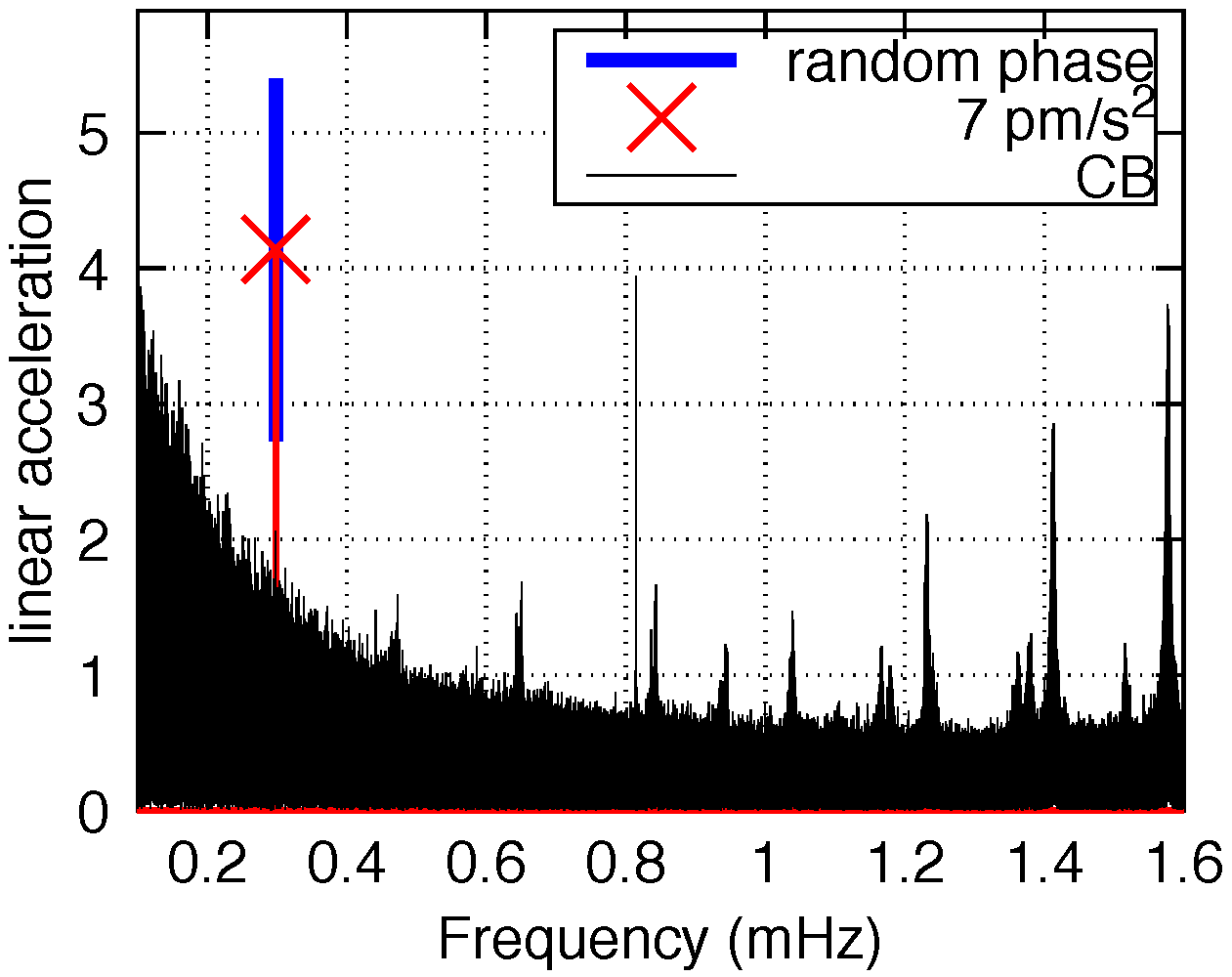}
 \caption{(Color online) 
Fourier ampliutude spectrum of pressure corrected gravity residuals at station CB (Canberra) similar to fig \ref{Fig2}. The dataset starts on 1 January 2010 and is 9 years long. The spectrum of one sinusoid at 0.3 mHz with the same data gaps and the same tapering as the gravity residuals is shown in red. Its amplitude is 7 pm/s$^2$. }
\label{Fig3}
\end{figure}

How much the detection level depends on the chosen admittance has been addressed in the electronic supplement \cite{Suplemental_info}. As long as an admittance between 
-2.5 and -4 nms$^{-2}$/hPa is used the gravity spectral level and hence the CDO detection level in Figs. \ref{Fig2} and \ref{Fig3} changes by less than 5\%.

 Multi-year long recordings cannot be analyzed without careful handling of artifacts in the data: times when the instrument behaved non-linearly due to saturation from large quakes, operator interference (Helium refills, cold head replacements, etc.) or other malfunctions. The gravity recordings are dominated by the tidal signal which can be well predicted \citep{ETERNA}. So we subtract from the data a synthetic tidal model for the stations that includes the effect of ocean loading. Subsequently the residual signal is visually inspected and we interactively flag large segments of bad data while short disturbances ($<$ 1 hr) are replaced by linear interpolation. We start with raw acceleration data sampled at 1 second intervals $\Delta g(t)$ \cite{gravimeter_data} and after processing arrive at band-passed (7200-300 s) data sampled at 4 minute intervals. The SG gravimeter data is calibrated by comparison against a co-located free-fall FG-5 absolute gravimeter in which a He-Ne laser and a rubidium clock provide atomic length and frequency standards, respectively. Since we are interested in the frequency band which is also occupied by the lowest order seismic free oscillations we have additionally flagged the hours and days following the largest quakes. These quakes are rare but would still lead to undesirable modal peaks in our frequency band. After this preprocessing of the data we arrive at a time series with $~$3\% flagged data. The flagged segments are zeroed before subsequent spectral analysis. 
 The Canberra results are similar to the BFO results however the noise is about two times larger.  Therefore we focus on the BFO results.        

In Figs. \ref{Fig2} and \ref{Fig3} we show Fourier amplitude spectra of the pressure corrected and Hanning tapered gravity residuals for BFO and CB respectively. A number of background signals are visible at both stations. A narrow, large amplitude spectral peak is seen near 0.8 mHz.  This is the fundamental monopole ($\ell=0$) free oscillation $_0S_0$ (or breathing mode) of the Earth.  This mode has a high $Q$ factor ($Q=5500$) and can remain excited for several months after a large earthquake.  Next to $_0S_0$ fundamental spheroidal free oscillation modes of the Earth ($_0S_\ell$) with angular order from $\ell=3$ to 9  are seen near 0.46, 0.64, 0.84, 1.03, 1.23, 1.41, and 1.57 mHz.  These modes are excited by large quakes. 

\begin{figure}[ht]
\smallskip
\includegraphics[width=1.0\columnwidth]{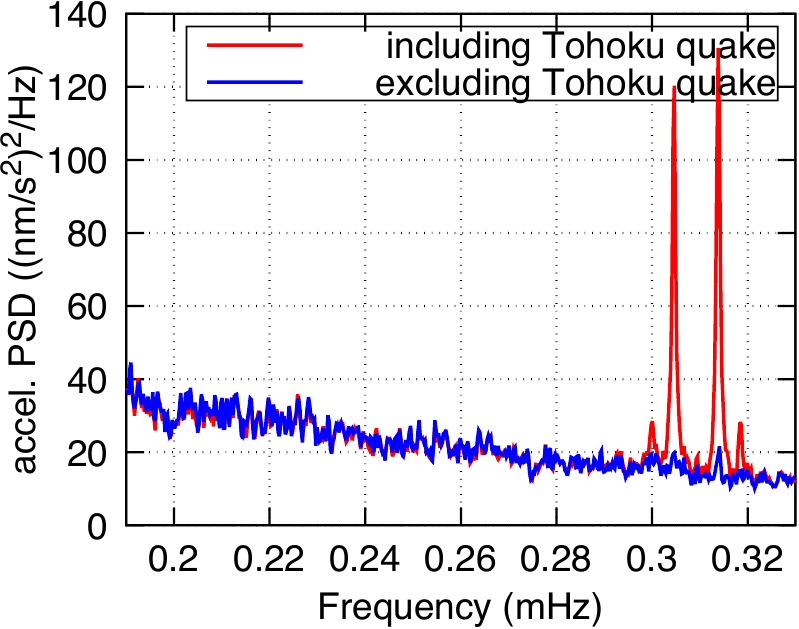}
 \caption{(Color online) 
Power spectral density of pressure corrected gravity residuals at BFO
for  the CDO target frequency band. The longer dataset includes the magnitude Mw9.1 Tohoku event (11 March 2011) and shows four out of the five singlets of the rotationally split fundamental spheroidal mode $_0S_2$. The longer dataset starts on 1. October 2010 and the shorter on 20. July 2011. They are 7.5 and 6.7 years long, respectively. }
\label{Fig4}
\end{figure}

The $\ell=2$ mode $_0S_2$ deserves special consideration because its frequency is very close to that expected for a CDO.  Figure \ref{Fig4} shows this mode clearly excited by the large Tohoku earthquake.  However, if we remove data for short periods after earthquakes then the mode is not a significant background.

To detect a phase coherent monochromatic signal in the gravity residues we use a Fourier amplitude spectrum of the full length dataset. The Fourier transform is optimal for this task because its basis functions are our target signal (matched filter). Furthermore we can rely on the fact that the Fourier amplitude of the phase coherent signal increases with the record length $N$, while the Fourier amplitude of the incoherent background only increases with $\sqrt{N}$. Thus the signal-to-noise ratio increases with the square root of the record length. 

In Figs. \ref{Fig2} and \ref{Fig3} we have also included the spectrum of a harmonic signal of known time-domain amplitude and identical gap structure and tapering as the gravity residues. If we inject the harmonic into the gravity residues by adding the two signals in the time domain, we expect a peak in the spectrum at the frequency of the injected harmonic whose amplitude depends on whether the two signals interfere constructively or destructively. Since the phase of the CDO signal is unknown we varied the initial phase of the injected harmonic in steps of 15$^{\circ}$ and tracked the variation of the peak amplitude. This variation is indicated with the blue vertical bar. Thus a 3 pm/s$^2$ harmonic gravity signal in the BFO gravity residues can show up in the Fourier amplitude spectrum with any value indicated by the blue bar.

Figure \ref{Fig2} shows that the total background at 0.3 mHz is significantly less than the 3 pm/s$^2$ calibration signal.  Therefore, we set an upper limit at this frequency of
\begin{equation}
\Delta g(0.3 {\rm mHz}) < 3\ {\rm pm/s}^2\, .
\label{2pmlimit}
\end{equation}
and a slightly larger value at 0.2 mHz. 
We now use this limit and Eq. \ref{Eq.dgfull} to set a limit on the product of CDO mass $m_D$ and orbital radius $a$.  The weakest limit is for CDO orbit inclination angle $\Theta_I$ that minimizes $F(\Theta_L)$ for a gravimeter at Latitude $\Theta_L$,
\begin{equation}
    F(\Theta_L)={\rm Min\, }_{\Theta_I}\, [\, {\rm Max}\, [ \delta_1,\delta_2,\delta_3]]\, .
    \label{Eq.F}
\end{equation}
For the Black Forest Observatory at $\Theta_L=48.33^\circ$ we have $F(48.33^\circ)\approx0.555$.  This occurs for $\Theta_I\approx 48^\circ$.  Using Eqs. \ref{Eq.dgfull},\ref{Eq.F} we have
$(m_D a)/(M_\oplus R_\oplus) < \Delta g(0.3{\rm mHz})/[g(2+\frac{\rho_c}{\bar\rho})F(48.33^\circ)].
$
Or using Eq. \ref{2pmlimit} our final limit is,
\begin{equation}
\frac{m_D a}{M_\oplus R_\oplus}<\frac{3\, {\rm pm/s^2}}{9.8\, {\rm m/s^2}\times 4.38\times 0.555} = 1.2\times 10^{-13}\, .   
\end{equation}
This Eq. is our main result.  We are able to set a very strict limit because the gravimeter is very sensitive.

In general we don't know the orbital radius $a$.  For reference let us consider $a\approx 0.1R_\oplus$.  Our limit is now,
\begin{equation}
    m_D<1.2\times 10^{-12}M_\oplus =4\times 10^{-18}M_\odot= 7\times 10^{12} \ {\rm kg}.
    \label{Eq.mdlimit}
\end{equation}
Of course, if $a$ is much smaller than $0.1R_\oplus$, the limit on $m_D$ becomes larger but this Eq. provides an order of magnitude expectation.  
Equation \ref{Eq.mdlimit} is over a million times smaller than the $10^{-11}M_\odot$ lowest mass probed by microlensing \cite{microlensing2019}.  

Since we don't observe any objects, we can set a limit on the probability of capture of a CDO in a collision with Earth.  If all of dark matter is made of CDOs (of a given mass) and CDOs orbit for a long time inside the Earth, then the probability of capture must be less than $\approx 10^{-3}$ for $m_D=10^{-18}M_\odot$ to less than $\approx 10^{-1}$ for $m_D=10^{-16}M_\odot$.  Please see the supplemental information for details \cite{Suplemental_info}.

One can search for CDOs in other solar system bodies.  The moon has no atmosphere and therefore little noise from atmospheric fluctuations.  The Lunar Surface Gravimeter was deployed on the moon during the Apollo 17 mission \cite{LunarSG,doi:10.1002/2014JE004724}.  Unfortunately, this instrument had a design flaw.  The space based gravitational wave detector LISA should be sensitive to GW radiation from CDOs moving in the sun or Jupiter, although a detection will likely require a significantly more massive object  \cite{Cornish:2018dyw}. %Note that the ground based detectors LIGO and VIRGO are only sensitive to GW frequencies above about 10 Hz.

In conclusion, dark matter could be composed of compact dark objects (CDOs). These objects may interact very  weakly  with  normal  matter  and  could  move freely inside the Earth.   We have searched superconducting gravimeter data and rule out CDOs moving in the Earth unless their masses $m_D$ and or orbital radii $a$ are very small so that $m_D\, a< 1.2\times 10^{-13} M_\oplus R_\oplus$.  Here $M_\oplus$ and $R_\oplus$ are the mass and radius of the Earth.  %

\acknowledgements{We acknowledge helpful discussions with Tamara Bogdanovic, Matt Caplan, Nicole Kinsman, Rafael Lang, Cole Miller, Maria Alessandra Papa and Walter Z\"urn. CJH thanks the Max Planck Institute for Gravitational Physics in Hannover and the KITP in Santa Barbara for their hospitality. CJH is supported in part by US Department of Energy grants DE-FG02-87ER40365 and DE-SC0018083. We gratefully acknowledge the work of the operators of the superconducting gravimeters: Harry McQueen and Tadahiro Sato from the National Astronomical Observatory, Mizusawa for the Canberra station and Thomas Forbriger and Peter Duffner from the Karlsruhe Institute of Technology for the Black Forest Observatory. We thank the data centers for archiving and freely distributing the gravimetric data and Th. Forbriger for porting the ETERNA software package for tidal predictions to UNIX \citep{ETERNA_UNIX}.
}

\bibliographystyle{apsrev}

\section{Appendix: Supplemental Information} 
In this appendix we first show the original gravimeter time series data.
Then we explore the efficiency of the barometric correction as a function of the chosen gravity-pressure admittance.
Next we calculate the gravimeter signal for a general CDO circular orbit and finally we calculate the advance of the perigee(s) in a CDO orbit.

\subsection{Time series data}

\begin{figure}[ht]
\smallskip
\includegraphics[width=1.0\columnwidth]{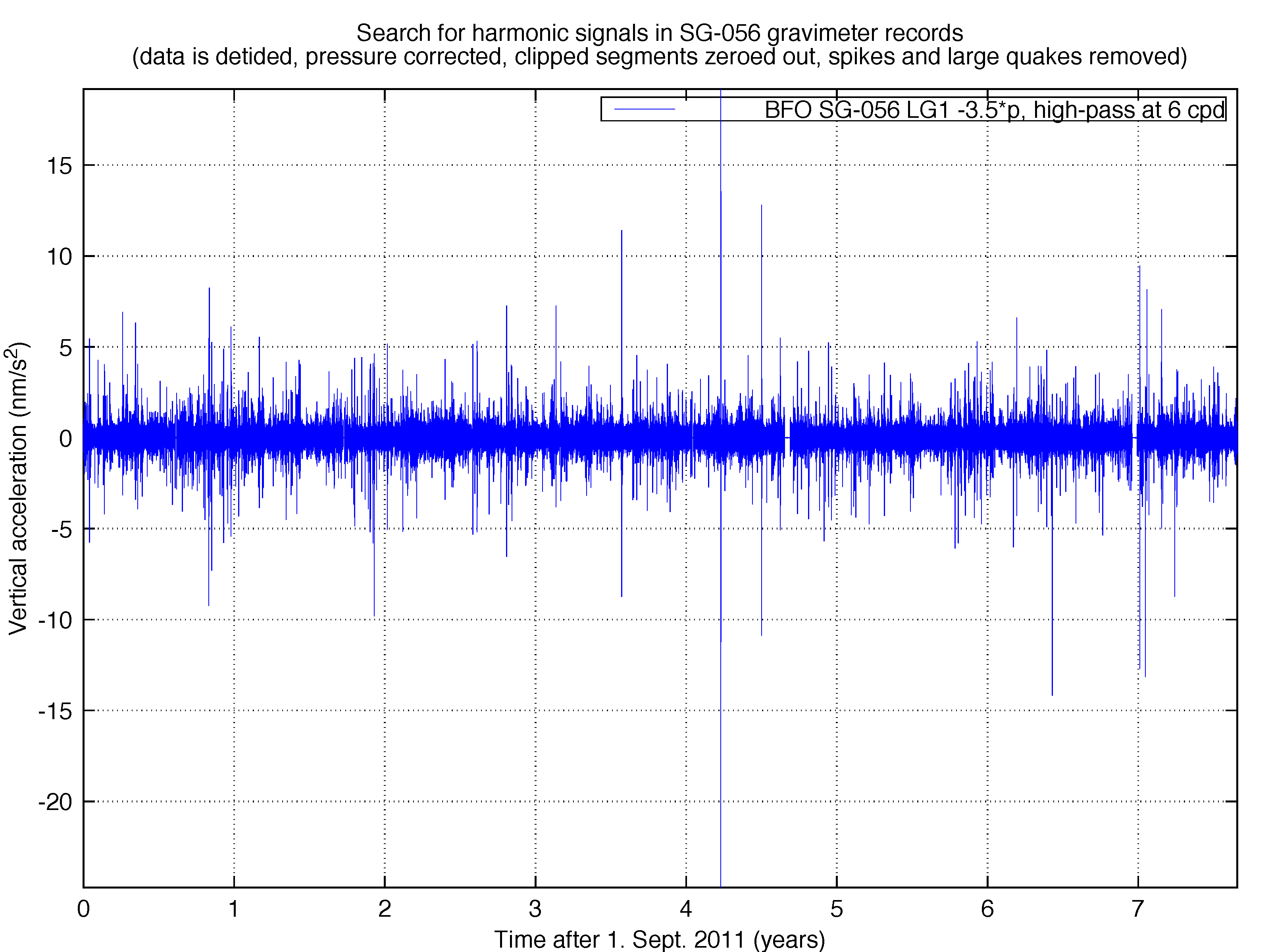}
 \caption{ Time series gravimeter data $\Delta g(t)$ from the Black Forest Observatory (BFO) for about eight years starting in Sep. 2011.}
\label{Fig1BFO}
\end{figure}

\begin{figure}[ht]
\smallskip
\includegraphics[width=1.0\columnwidth]{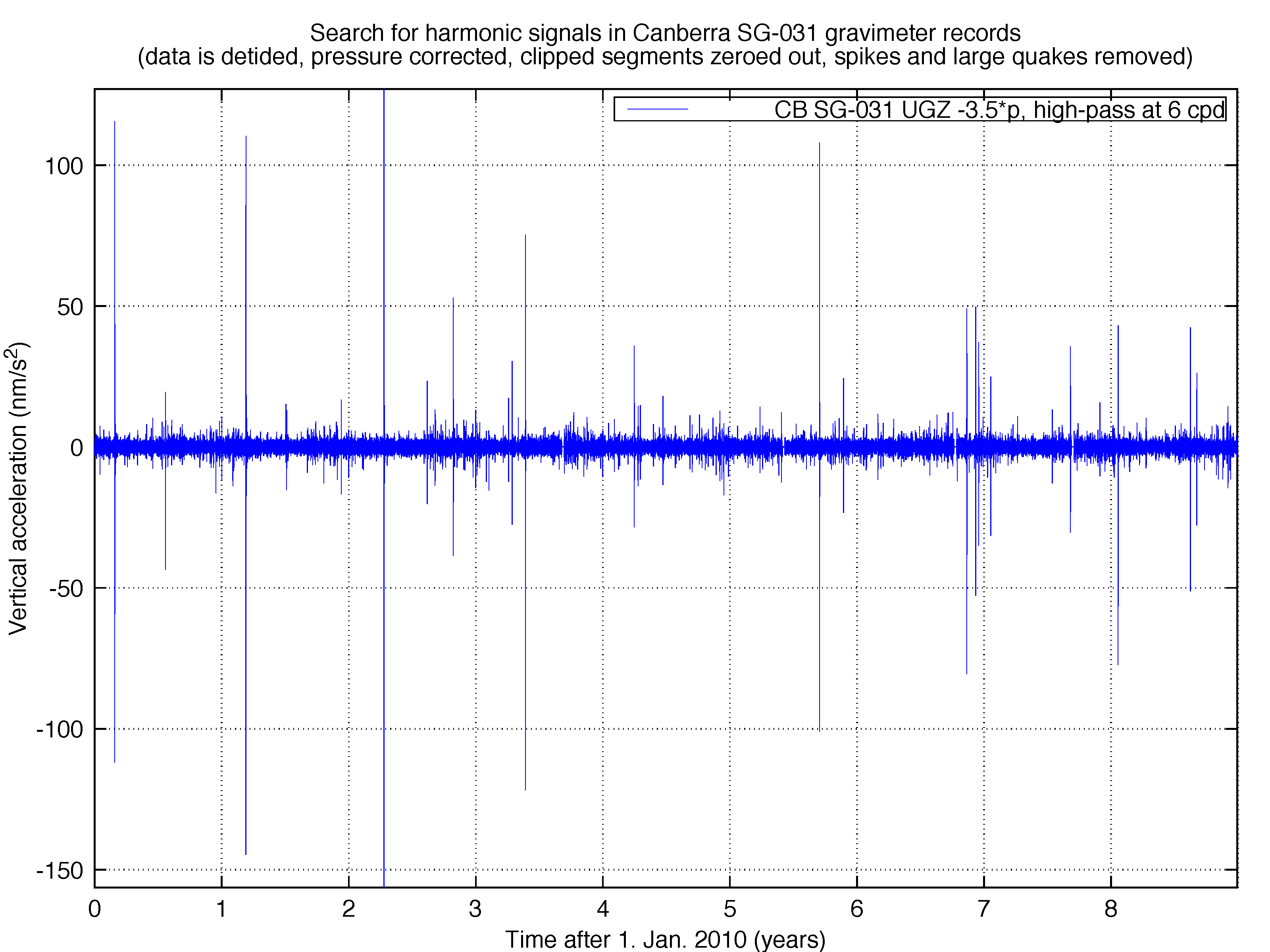}
 \caption{ Time series gravimeter data $\Delta g(t)$ from the Canberra station for nine years starting in Jan. 2010. More of the Earth quake signal is left in this dataset compared to BFO (fig.\ref{Fig1BFO}) and hence more low-frequency modes can be seen in the spectrum of fig. \ref{Fig3}. 
 }.
\label{Fig1CO}
\end{figure}

Here we show the edited time series data of gravitational residues for the Black Forest Observatory (BFO) in Fig. \ref{Fig1BFO} and for the Canberra station in Fig. \ref{Fig1CO}. The signal from the Earth tides have already been removed as well as any clipped data segments, large quakes and the instrumental drift.  

\subsection{Barometric pressure correction}

\begin{figure}[ht]
\smallskip
\includegraphics[width=1.0\columnwidth]{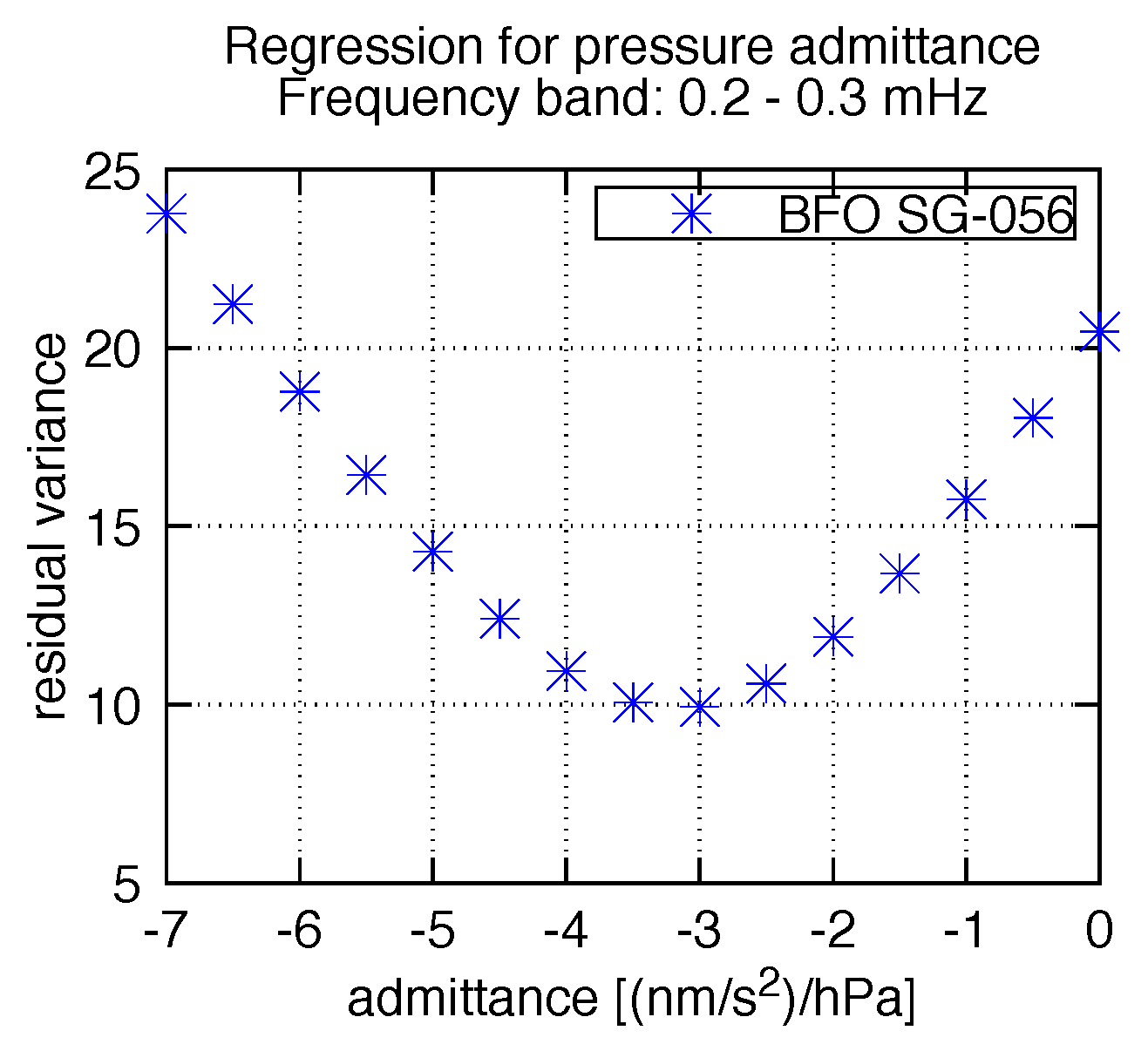}
 \caption{Residual variance in pressure corrected gravity spectra in the band 0.2 - 0.3 mHz.
 }
\label{Fig_Parabola}
\end{figure}

To assess the influence of the chosen barometric correction on the detection level in the Fourier gravity spectra (Figs \ref{Fig2} and \ref{Fig3}) we vary the pressure admittance 
between 0 and -7 nms$^{-2}$/hPa in steps of 0.5 nms$^{-2}$/hPa 
(Fig. \ref{Fig_Parabola}). 
The admittance is then used to scale the barogram before subtracting it in the time domain from the gravity record. After FFT the residual variance is evaluated in the band 0.2 - 0.3 mHz.  A broad minimum between -2.5 and -4 nms$^{-2}$/hPa is evident where the variance varies by 10\% only. 

\subsection{General Orbit}
Consider a CDO in a circular orbit of radius $a$ that is inclined by an angle $\Theta_0$ w.r.t. the plane of the equator.  Let the gravimeter be at Latitude $\Theta_L$.  Choose a coordinate system fixed in space with the $z$ axis along the Earth's rotation axis and the $x$ axis along the intersection of the CDO orbital plane and the plane of the equator.

The coordinates of the gravimeter as the Earth rotates are,
\begin{equation*}
    X_g=R_\oplus \cos\Theta_L \cos( \omega_\oplus t),
    \end{equation*}
    \begin{equation*}
    Y_g=R_\oplus\cos\Theta_L \sin (\omega_\oplus t),
    \end{equation*}
    \begin{equation}
        Z_g=R_\oplus \sin\Theta_L\, .
    \end{equation}
Here $\omega_\oplus=2\pi/({\rm 1\ day})$.  The coordinates of the CDO in its orbital plane are $X'=a\cos(\omega_r t)$, $Y'=a\sin(\omega_r t)$, $Z'=0$.  Rotating this coordinate system through an angle $\Theta_0$ about the $X'=X$ axis yields the coordinates of the COD in the space fixed frame,
\begin{equation*}
    X=a\cos(\omega_r t),
    \end{equation*}
    \begin{equation*}
    Y=a\cos\Theta_L\sin(\omega_r t),
    \end{equation*}
    \begin{equation}
    Z=A\sin\Theta_L\sin(\omega_r t)\, .
\end{equation}
Here $\omega_r=2\pi\nu(a)$, see Eq. \ref{nu_r}.  We now calculate the square of the distance between the CDO and the gravimeter,
\begin{equation}
    (X-X_g)^2+(Y-Y_g)^2+(Z-Z_g)^2=R_\oplus^2+a^2-2R_\oplus a \Delta(t)\, .
\end{equation}
Here the quantity $\Delta$ can be written,
\begin{equation}
\Delta(t)=\Delta_1+\Delta_2+\Delta_3,
\end{equation}
with,
\begin{equation}
 \Delta_1=\cos\Theta_L\cos^2\frac{\Theta_0}{2}\cos[(\omega_r-\omega_\oplus)t]\, ,  
\end{equation}
\begin{equation}
 \Delta_2=\cos\Theta_L\sin^2\frac{\Theta_0}{2}\cos[(\omega_r+\omega_\oplus)t]\, ,  
\end{equation}
\begin{equation}
\Delta_3=\sin\Theta_L\sin\Theta_0\sin(\omega_r t)\, .
\end{equation}
Following similar steps as the example in the text and assuming $a\ll R_\oplus$, the time dependent gravimeter signal is,
\begin{equation}
\frac{\Delta g(t)}{g}=\bigr(\frac{\bar\rho(a)}{\bar\rho(R_\oplus)}+2\bigr)\bigl(\frac{m_D\, a}{M_\oplus R_\oplus}\bigr)\Delta(t)\, .
\end{equation}
We see that in general there is a gravimeter signal at angular frequency $\omega_r$ (coming from the $\Delta_3$ part) and at the two rotational side bands $\omega_r\pm\omega_\oplus$ (from the $\Delta_1$ and $\Delta_2$ parts).

\subsection{Advance in the perigee of CDO orbits}

If the density is constant, the general CDO orbit is a closed ellipse with the center of the Earth at the center of the ellipse (not at one of the foci).  If the density changes with radius (but is still assumed to be spherically symmetric) the elliptical orbit remains in one plane but it no longer closes.  Instead, the orientation of the major axis of the ellipse will advance with time.  This is not unlike the advance of the perihelion of Mercury.  Note that for CDO orbits there are two points in the orbit where the distance to the center of the Earth is minimum (perigee) and two points where the distance is maximum (apogee).  Thus we can talk about the advance in the perigee (either one) of the CDO's orbit.

For simplicity we consider nearly circular orbits in the inner core of the Earth.  The solid inner core extends to $R_{ic}=0.19173R_\oplus=1221.5$ km  \cite{DZIEWONSKI1981297}.  It is possible that dynamical friction reduces the radius of CDO orbits so that they may spend a lot of time orbiting in the inner core.  Furthermore, dynamical friction may tend to circularize CDO orbits so that the remaining eccentricity is small.  In general larger radius orbits will give larger signals and may be easier to rule out.  Therefore we will focus here on smaller radius orbits in the inner core.

The density of the inner core $\rho(r)$ is approximately 
\begin{equation}
\rho(r)=\rho_c-\rho_1 \frac{r^2}{R_\oplus^2}\, ,
\end{equation}
valid for $r<R_{ic}$.  Here $\rho_c=13.0885$ g/cm$^3$ and $\rho_1=8.8381$ g/cm$^3$ \cite{DZIEWONSKI1981297}.  The enclosed mass is $M(r)=\int^r4\pi r'^2dr'\rho(r')$ and the average density is $\bar\rho(r)=M(r)/(4\pi r^3/3)$,
\begin{equation}
\bar\rho(r)=\rho_c-\rho_1\frac{3r^2}{5R_\oplus^2}\, .
\end{equation}
The orbital frequency $\nu$ then follows from Eq. \ref{omega},
\begin{equation}
\nu=\nu_0\bigl(1-\frac{3\rho_1 r^2}{5\rho_c R_\oplus^2}\bigr)^{1/2}\, ,
\label{nu_r}
\end{equation}
with $\nu_0=(G\rho_c/3\pi)^{1/2}$.  The angular period $T=1/\nu$ is the time it takes for the angle $\phi$ in polar coordinates to advance by $2\pi$.  Instead, the radial period $T_r$ is the time it takes for $r$ to execute one small amplitude oscillation about the equilibrium radius of a circular orbit.  This can be easily found by expanding the effective radial potential for small oscillations.  We have for the radial frequency $\nu_r=1/T_r$,
\begin{equation}
\nu_r= 2\nu_0\bigl(1-\frac{9\rho_1 r^2}{10\rho_c R_\oplus^2}\bigl)^{1/2}\, .
\end{equation}
In general $\nu_r\approx 2\nu$ because for an ellipse centered on the origin, the radius undergoes two complete oscillation periods as $\phi$ goes around once (there are two perigee per orbit).

The fractional advance of a perigee per orbit $\Delta$ (in units of $360^\circ$) is,
\begin{equation}
\Delta=\frac{\nu-\frac{1}{2}\nu_r}{\nu}\approx \frac{3\rho_1 r^2}{20\rho_c R_\oplus^2}\, .
\end{equation}
We have $\Delta\le 0.00372$ for $r\le R_{ic}$.  The time it takes for a perigee to advance through 360$^\circ$ is,
\begin{equation}
T_a=\frac{1}{\nu \Delta}   \ge 10.4 \ {\rm days}\, ,
\end{equation}
for $r\le R_{ic}$.

In general, the original gravimeter signal at the frequency $\nu$ will be modulated, as the perigee advances, at the significantly lower frequency $\nu\Delta$. 

\subsection{Collision rate and capture probabilities}

If all of dark matter is made of CDOs of mass $m_D$, the CDO density is
\begin{equation}
    n_D = 3\times 10^{-34} {\rm m^{-3}}\bigl(\frac{10^{-18}M_\odot}{m_D}\bigr)\, .
\end{equation}
We assume a geometric cross section for colliding with Earth 
\begin{equation}
    \sigma=\pi R_\oplus^2\, .
\end{equation}
The total number of Earth CDO collisions $N_{\rm coll}$ in a time $\tau$ is
\begin{equation}
    N_{\rm coll}=\sigma\, v\, \tau\, n_D\, .
\end{equation}
Here the velocity of the CDOs relative to Earth is assumed to be v=220 km/s.  The maximum value for $\tau$ is the age of the Earth $5\times 10^9$ y.  This gives
\begin{equation}
    N_{\rm coll} = 1300\ \bigl(\frac{10^{-18}M_\odot}{m_D}\bigr)\bigl(\frac{\tau}{5\times 10^9{\rm y}}\bigr)\, .
\end{equation}
Since we do not observe any captured objects, the probability of capture must be less than of order $1/N_{\rm coll}$.  For large $\tau$ the capture probability must be smaller than $\approx 10^{-3}$ for $m_D=10^{-18}M_\odot$ to smaller than $\approx 10^{-1}$ for $m_D=10^{-16}M_\odot$.
\end{document}